\documentclass[prd,aps,groupedaddress,10pt]{revtex4}
\usepackage{epsf,epsfig,graphicx}

  \newcount\hour \newcount\minute
  \hour=\time \divide \hour by 60
  \minute=\time
  \newcommand{\mydate}{\ \today \ - \number\hour :\ifnum \minute<10 0\fi
\number\minute}

\def\nslash{n\!\!\!\slash}
\def\bnslash{\bar n\!\!\!\slash}
\def\pslash{p\!\!\!\slash}
\def\ks{K^*_0}

\def\km{K^{*-}_0}
\def\k0{{K^{*0}_0}}
\def\bk0{\overline{K^{*}_0}^0}

\newcommand{\bea}{\begin{eqnarray}}
\newcommand{\eea}{\end{eqnarray}}

\begin{document}

\title{Study of $K^*_0(1430)$ and $a_0(980)$ from $B\to K^*_0(1430)\pi$ and $B\to a_0(980)K$ Decays }

\author{Yue-Long Shen$^b$,
Wei Wang$^b$\footnote{Email:wwang@mail.ihep.ac.cn}, Jin Zhu$^b$
and Cai-Dian L\"u$^{a,b}$} \affiliation{\it \small
 a CCAST (World Laboratory), P.O. Box 8730,
   Beijing 100080, P.R. China\\
\it \small  b Institute of High Energy Physics, CAS, P.O.Box
918(4), \it \small 100049, P.R. China\footnote {Mailing address}}

\begin{abstract}
We use the decay modes $B \to K^*_0(1430) \pi$ and $B \to a_0(980)
K$ to study the scalar mesons  $K^*_0(1430)$ and $a_0(980)$ within
perturbative QCD framework. For $B \to K^*_0(1430) \pi$, we
perform our calculation in two scenarios of the scalar meson
spectrum. The results indicate that scenario II is more favored by
experimental data than scenario I. The important contribution from
annihilation diagrams can enhance the branching ratios about
$50\%$ in scenario I, and about $30\%$ in scenario II. The direct
$CP$ asymmetries in $B \to K^*_0(1430) \pi$ are small, which are
consistent with the present experiments. The predicted branching
ratio of $B \to a_0(980) K$ in scenario I differs from the
experiments by a factor 2, which indicates $a_0(980)$ can not be
interpreted as $\bar qq$.

\end{abstract}

\maketitle 

\section{introduction}

The scalar meson spectrum is an interesting topic for both
experimental and theoretical studies, but the underlying structure
of the light scalar mesons is still under controversy. Many scalar
meson states have been found in experiments: isoscalar states
$\sigma(600)$, $f_0(980)$, $f_0(1500)$, $f_0(1370)$, $f_0(1710)$;
the isovector states $ a_0(980)$, $a_0(1450)$ and isodoublets
$\kappa(800)$, $K^*_0(1430)$.  In the literature, there are many
schemes for the classification of these states
\cite{scenarioI,scalarb,scalarc,scalar}. Here are two typical
scenarios: the members of the lower mass nonet $\sigma(600)$,
$\kappa(800)$, $f_0(980)$, and $a_0(980)$ are treated as the
lowest lying $q\bar q$ states, while $K^*_0(1430)$ {\it et} {\it
al}. which form the higher mass nonet are the first excited $q\bar
q$ states; In scenario II, the members of the lower mass nonet are
treated as the four-quark states \cite{scenarioI}. Then the higher
mass nonet is considered as the lowest lying $\bar qq$ states.
There are also other schemes to classify these states, for
example, $\sigma(600)$ and $\kappa(800)$ are not considered as the
physical states, $a_0(980)$ (or $a_0(1450)$), $f_0(980)$,
$K^*_0(1430)$ and $f_0(1500)$ form the $q\bar{q}$ nonet
\cite{scalarc}. In this paper, we study the scalar mesons in the
first two scenarios.

Although intensive study has been given to the decay property of
the scalar mesons, the production of these mesons can provide a
different unique insight to the mysterious structure of these
mesons.  Compared with $D$ meson decays, the phase space in $B$
decays is larger, thus $B$ decays can provide a better place to
study the scalar resonances. Experimentally, $B$ meson decay
channels with a final state scalar meson have been measured in $B$
factories for several years \cite{first}. Much more measurements
have been reported by BaBar and Belle
\cite{garmash,babar,belle,abe,new} recently (see \cite{hfag} for
more experimental data). On the theoretical side, the $B$ decays
which involve a scalar meson have been systematically studied
using QCD factorization (QCDF) approach by Cheng, Chua and Yang
\cite{cheng}. They draw the conclusion that scenario II is more
preferable.
For example, in scenario I, the predicted branching ratio of
$B^-\to \overline{K^*_0}^0\pi^-$ without annihilation is only
about $1\times 10^{-5} $, which is much smaller than the
experimental results. In order to explain the large data, the
annihilation contribution is required to be large. But in this
scenario, $a_0(980)$ is also a $q\bar q$ state and the $SU(3)$
symmetry implies a much large annihilation contribution to $B\to
a_0(980)K$. This large annihilation contribution can lead to a
larger branching ratio than the experimental upper bound. Then it
is concluded that scenario I is less preferable than scenario II.

The annihilation topology contribution plays such an important
role that we should pay much more attention to it. In QCD
factorization approach, which is based on collinear factorization,
the so-called endpoint singularity makes the annihilation
contribution divergent. This might be resolved by applying the
``zero-bin" subtractions  which will lead to new factorization
theorems in rapidity space \cite{zerobin}. The more popular way to
handle this divergence is the Perturbative QCD (PQCD) approach
\cite{kls,lucd}. Using this approach, some pure annihilation type
decays have been studied and the results are consistent with the
experiments \cite {luanni}, which indicates it a reliable method
to deal with annihilation diagrams. In the present paper, we will
use the PQCD approach to calculate the decay modes $B\to
K^*_0(1430)\pi$ and $B\to a_0(980)K$ (in the following, we use
$\ks$ and $a_0$ to denote $\ks(1430)$ and $a_0(980)$ for
convenience). We will examine how large the annihilation topology
contribution is, and find whether there are enough reasons to
determine which scenario is more appropriate through these decay
channels.

This paper is organized as follows: in the next section, we give a
brief review of the scalar meson wave functions, which are
important inputs in PQCD approach. What followed is the analysis
of the $B\to K^*_0\pi$ and $B\to a_0K$ decays. The numerical
results and the discussions are given in the section \ref{sec4}.
Our conclusions are presented in the final part.


\section{Scalar meson distribution amplitudes}

In the two-quark picture, the decay constants $f_S$ and $\bar f_S$
for a scalar meson $S$ are defined by:
\begin{eqnarray}
\langle S(p)|\bar{q}_2 \gamma_{\mu}q_1 |0
\rangle=f_Sp_{\mu},\,\,\, \langle S(p)|\bar{q}_2 q_1 |0
\rangle=\bar{f}_S m_S,
\end{eqnarray} where $m_S(p)$ is the mass (momentum) of the scalar
meson. The vector current decay constant $f_S$ is zero for neutral
scalar mesons $\sigma$, $f_0$ and $a_0^0$ due to the charge
conjugation invariance or G parity conservation. In the $SU(3)$
limit, the vector decay constant of $K^*_0$ also vanishes. After
including the $SU(3)$ symmetry breaking, it only gets a very small
value which is proportional to the mass difference of the
constituent quarks. The scalar density decay constant $\bar{f}_S$
can be related to the vector one by the equation of motion:   \bea
\mu_Sf_S=\bar{f}_S, \eea where $\mu_S$ is defined by
$\mu_S=\frac{m_S}{m_{q_2}(\mu)-m_{q_1}(\mu)}$, which is scale
dependent, thus the  scalar decay constant is also scale
dependent. Many model calculations have been performed for the
$\ks$ and $a_0$ \cite{decayconstant}. In this paper we will use
the values from QCD sum rules in ref. \cite{cheng}. Fixing the
scale at $1\mbox{GeV}$, we specify them below:
\begin{eqnarray}
\nonumber
    \mbox{scenario} \,{\rm I}:\,\,\,\,
   & \bar{f}_{\ks}=-(300\pm
30)\mbox{MeV},&f_{\ks}=-(25\pm 2)\mbox {MeV} ,\\
&\bar{f}_{a_0}=(365\pm 20)\mbox{MeV};
\\
\mbox{scenario} \,{\rm II}:\,\,\,\,&\bar{f}_{\ks}=(445\pm
50)\mbox{MeV}, & ~f_{\ks}=(37\pm 4)\mbox{MeV}.
 \end{eqnarray}

Now we turn to the distribution amplitudes.  The scalar meson's
light-cone distribution amplitude is defined by:
 \bea \langle
S(p)|\bar{q}_{1\beta}(z)q_{2\alpha}(0)|0\rangle \nonumber
&=&\frac{1}{\sqrt{6}}\int^1_0dxe^{ixp \cdot z}\bigg\{ \pslash
\phi_S(x)+ m_S\phi^S_S(x)-\frac{1}{6}
m_S\sigma_{\mu\nu}p^{\mu}z^{\nu}\phi^{\sigma}_S(x)\bigg\}_{\alpha\beta}\\
&=& \frac{1}{\sqrt{6}}\int^1_0dxe^{ixp \cdot z}\bigg\{ \pslash
\phi_S(x)+ m_S\phi^S_S(x)+
m_S(\nslash\bnslash-1)\phi^T_S(x)\bigg\}_{\alpha\beta}, \eea
 where $n=(1,0,0_T)$ and $\bar{n}=(0,1,0_T)$ are dimensionless vectors on
the light cone, and $n$ is parallel with the moving direction of
the scalar meson.
 The distribution amplitudes $\phi_S(x)$,
$\phi^S_S(x)$ and $\phi^{\sigma}_S(x)$ are normalized as: \bea
\int^1_0dx\phi_S(x)=\frac{f_S}{2\sqrt{6}},\,\,\,\,\,\,
\int^1_0dx\phi^S_S(x)=\int^1_0dx\phi^{\sigma}_S(x)=\frac{\bar{f}_S}{2\sqrt{6}},
\eea and $\phi^T_S(x)=\frac{1}{6}\frac{d}{dx}\phi^{\sigma}_S(x)$.

In general, the twist-2 light cone distribution amplitude
$\phi_S(x)$ can be expanded as:
 \bea \phi_S(x,\mu)\nonumber
&=&\frac{\bar{f}_S(\mu)}{2\sqrt{6}}6x(1-x)\bigg[B_0(\mu)
+\sum\limits^{\infty}_{m=1}B_m(\mu)C_m^{3/2}(2x-1)\bigg]\\
&=&\frac{{f}_S(\mu)}{2\sqrt{6}}6x(1-x)\bigg[1+\mu_S\sum\limits^{\infty}_{m=1}
B_m(\mu)C_m^{3/2}(2x-1)\bigg], \eea where $B_m(\mu)$ and
$C_m^{3/2} (x)$ are the Gegenbauer moments and Gegenbauer
polynomials respectively. The Gegenbauer moments $B_1$, $B_3$ of
distribution amplitudes for $\ks $ and $a_0$ have been calculated
in \cite{cheng} as \bea \mbox{scenario} \,{\rm I}:\,\,\,\,
B_1=0.58\pm 0.07,\;\;\;\; B_3=-1.20\pm 0.08;\\ B_1=-0.93\pm
0.10,\;\;\;\; B_3=0.14\pm 0.08;\\ \mbox{scenario} \,{\rm
II}:\,\,\,\, B_1=-0.57\pm 0.13,\;\;\;\; B_3=-0.42\pm 0.22, \eea
where the second line is for $a_0$, and the others are for $\ks$.
These values are also taken at $\mu=\mbox{1GeV}$.

As for the twist-3 light-cone distribution amplitudes, there is no
study on their explicit Gegenbauer moments so far, so we take the
asymptotic form in our numerical calculation: \bea
\phi^S_S(x)=\frac{\bar{f}_S}{2\sqrt{6}},\,\,\,\,\,\phi^T_S(x)=\frac{\bar{f}_S}{2\sqrt{6}}(1-2x).
\eea

In our calculation, we will choose the momentum fraction on the
anti-quark, thus we should use $\phi_S(1-x)$,
$\phi^S_S(1-x)=\phi^S(x)$ and $\phi_S^T(1-x)=-\phi_S^T(x)$. But in
the amplitudes, for simplicity, we use $\phi_S(x)$ to denote
$\phi_S(1-x)$ . It is similar for the pseudoscalar meson.

\section{The perturbative QCD calculation}
\label{sec3}

In this section we will give the decay amplitudes for the $B \to
\ks \pi$ and $B \to a_0 K$ decays in the PQCD approach. The PQCD
approach is based on the $k_T$ factorization \cite{botts}, where
we keep the transverse momentum of the partons in a meson. In this
approach there is no divergence when the longitudinal parton
momentum fraction falls into the endpoint region. The weak
decay matrix element can be completely factorized \cite{review}:
\begin{eqnarray}
{\cal A}=\phi_B\otimes H^{(6)}
\otimes J \otimes S\otimes \phi_{M_1}\otimes \phi_{M_2},
\end{eqnarray}
 where $\phi_B$, $\phi_{M_1}$ and $\phi_{M_2}$ denote
the wave functions of the $B$ meson and the light mesons,
respectively. $S$ and $J$ denote the Sudakov form factor and the
jet function respectively. The Sudakov form factor comes from
$k_T$ resummation which kills the end-point singularities. The jet
function is from threshold resummation which can organize the
large double logarithms in the hard kernel. The symbol $\otimes$
denotes convolution of the parton longitudinal momentum fractions
and the transverse momentum. $H^{(6)}$  is the six-quark hard
scattering kernel, which consists of the effective four quark
operators and a hard gluon to connect the spectator quark in the
decay. The standard four-quark operators describing the $b\to s$
transition are defined as \cite{buras}:
\begin{itemize}
 \item  current--current tree operators
    \begin{eqnarray}
  O_{1}=({\bar{u}}_{\alpha}b_{\beta} )_{V-A}
               ({\bar{s}}_{\beta} u_{\alpha})_{V-A},
    \ \ \ \ \ \ \ \ \
  O_{2}=({\bar{u}}_{\alpha}b_{\alpha})_{V-A}
               ({\bar{s}}_{\beta} u_{\beta} )_{V-A},
    \label{operator02}
    \end{eqnarray}
     \item  QCD penguin operators
    \begin{eqnarray}
      O_{3}=({\bar{s}}_{\alpha}b_{\alpha})_{V-A}\sum\limits_{q^{\prime}}
           ({\bar{q}}^{\prime}_{\beta} q^{\prime}_{\beta} )_{V-A},
    \ \ \ \ \ \ \ \ \
    O_{4}=({\bar{s}}_{\beta} b_{\alpha})_{V-A}\sum\limits_{q^{\prime}}
           ({\bar{q}}^{\prime}_{\alpha}q^{\prime}_{\beta} )_{V-A},
    \label{eq:operator34} \\
     \!\!\!\! \!\!\!\! \!\!\!\! \!\!\!\! \!\!\!\! \!\!\!\!
    O_{5}=({\bar{s}}_{\alpha}b_{\alpha})_{V-A}\sum\limits_{q^{\prime}}
           ({\bar{q}}^{\prime}_{\beta} q^{\prime}_{\beta} )_{V+A},
    \ \ \ \ \ \ \ \ \
    O_{6}=({\bar{s}}_{\beta} b_{\alpha})_{V-A}\sum\limits_{q^{\prime}}
           ({\bar{q}}^{\prime}_{\alpha}q^{\prime}_{\beta} )_{V+A},
    \label{eq:operator56}
    \end{eqnarray}
 \item electro-weak penguin operators
    \begin{eqnarray}
     O_{7}=\frac{3}{2}({\bar{s}}_{\alpha}b_{\alpha})_{V-A}
           \sum\limits_{q^{\prime}}e_{q^{\prime}}
           ({\bar{q}}^{\prime}_{\beta} q^{\prime}_{\beta} )_{V+A},
    \ \ \ \
    O_{8}=\frac{3}{2}({\bar{s}}_{\beta} b_{\alpha})_{V-A}
           \sum\limits_{q^{\prime}}e_{q^{\prime}}
           ({\bar{q}}^{\prime}_{\alpha}q^{\prime}_{\beta} )_{V+A},
    \label{eq:operator78} \\
     O_{9}=\frac{3}{2}({\bar{s}}_{\alpha}b_{\alpha})_{V-A}
           \sum\limits_{q^{\prime}}e_{q^{\prime}}
           ({\bar{q}}^{\prime}_{\beta} q^{\prime}_{\beta} )_{V-A},
   \ \ \ \
  O_{10}=\frac{3}{2}({\bar{s}}_{\beta} b_{\alpha})_{V-A}
           \sum\limits_{q^{\prime}}e_{q^{\prime}}
           ({\bar{q}}^{\prime}_{\alpha}q^{\prime}_{\beta} )_{V-A},
    \label{operator9x}
    \end{eqnarray}

 \end{itemize}
where $q^\prime=(u,d,s,c,b)$. The 10 operators together with their
QCD-corrected Wilson coefficients form the effective Hamiltonian:
 \begin{eqnarray}
 {\cal H}_{eff} &=& \frac{G_{F}}{\sqrt{2}}
    \left \{ V_{ub} V_{us}^*  \Big[
     C_{1}({\mu}) O_{1}({\mu})
  +  C_{2}({\mu}) O_{2}({\mu})\Big]
  -V_{tb}V_{ts}^* {\sum\limits_{i=3}^{10}} C_{i}({\mu}) O_{i}({\mu})\right\}
 .
 \label{eq:hamiltonian01}
 \end{eqnarray}

The partition of the perturbative and non-perturbative region
(factorization scale) is quite arbitrary, but the full amplitude
should be independent of the partition. We usually take the
largest virtuality of internal particles as the factorization
scale, which is of order $\sqrt{m_B\Lambda}$. The leading order
Wilson coefficients will be evolved to this scale. As the
factorization scale in PQCD approach is smaller than  the
factorization scale of QCDF, which is about $m_B$,  a large
enhancement of the Wilson coefficients occurs, especially for the
penguin operators \cite{kls,lucd}.

\subsection{$B \to \ks \pi$ decays}

\begin{figure}[tb]
\vspace{-0.4cm}
\begin{center}
\psfig{file=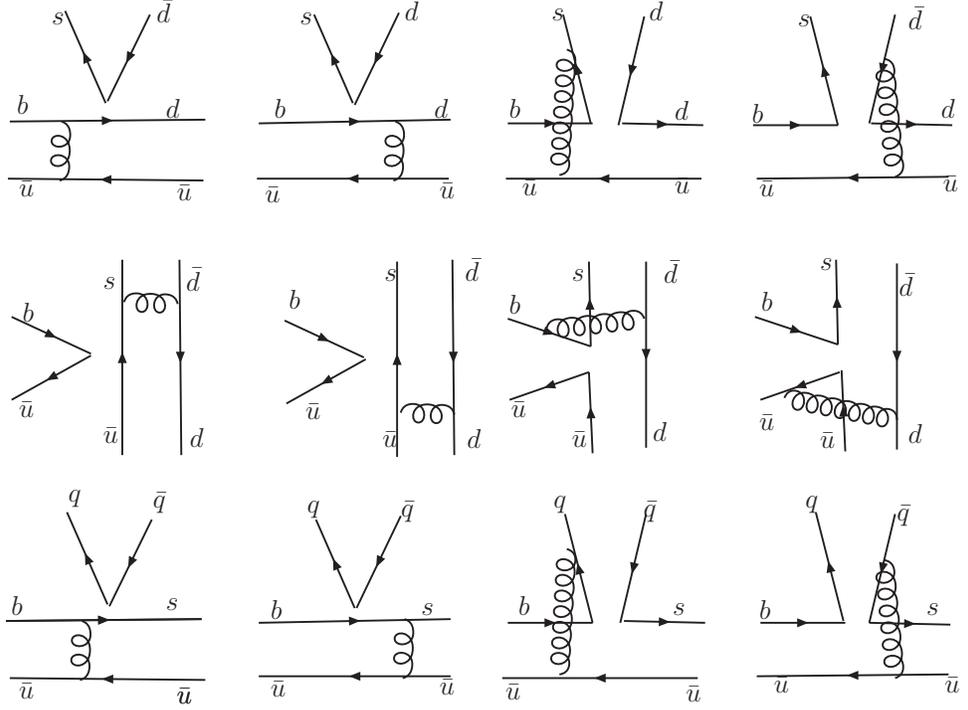,width=15.0cm,angle=0}
\end{center}
\vspace{-9.5cm} \caption{ The leading order Feynman diagrams in
PQCD for $B\to \ks \pi$ and $B \to a_0 K$}\label{fig1}
\end{figure}

The leading order Feynman diagrams for these decays in PQCD
approach are given in Fig.~\ref{fig1}. The decay amplitude for
each diagram can be obtained by contracting the hard scattering
kernels and the meson's wave functions. According to the power
counting in PQCD approach \cite{lipc}, the first two emission
diagrams in Fig.~\ref{fig1} give the dominant contribution. For
the $(V-A)(V-A)$ kind of operators, the decay amplitudes for these
two diagrams are given by:
\begin{eqnarray}
F^L_{B\to\pi}(a) &=& \frac{32 \pi}{ 3} m_B^4 f_{\ks}\int_0^1 dx_1
dx_3 \int_0^{\infty} b_1db_1\, b_3db_3\, \phi_B(x_1,b_1)
\bigg\{a(t) E_{e}(t)
\nonumber \\
& &\times  \left[ (1+x_3)\phi_{\pi}^A(x_3)+r_{\pi}(1-2x_3) \left(
\phi_{\pi}^P(x_3)+\phi_{\pi}^T(x_3) \right) \right]
h_{e}(x_1,x_3,b_1,b_3)
\nonumber\\
& &\;\;\;\;\;\; +2r_{\pi} \phi_{\pi}^P({x_3})a(t')E_{e}(t')
h_{e}(x_3,x_1,b_3,b_1) \bigg\} ,\label{me4}
\end{eqnarray}
where $r_\pi =m_0^\pi /m_B$, $m_0^\pi$ is the chiral enhancement
scale and $a$ is the corresponding Wilson coefficient. $E_e(t)$ is
defined as
\begin{eqnarray}
E_{e}(t)&=&\alpha_s(t) \exp[-S_B(t)-S_{\pi}(t)].
\end{eqnarray}
 For the $(V-A)(V+A)$ kind of
operators, the decay amplitudes for these two diagrams are given
by:
 \begin{eqnarray}
 F^R_{B\to\pi}(a) &=& -\frac{64 \pi}{ 3}
m_B^4r_{\ks} {\overline{f_{\ks}}}\int_0^1 dx_1 dx_3 \int_0^{\infty}
b_1db_1\, b_3db_3\, \phi_B(x_1,b_1)\bigg\{a(t) E_{e}(t)
\nonumber \\
& &\times   \left[ \phi_{\pi}^A(x_3)+r_{\pi} x_3 \left(
\phi_{\pi}^P(x_3) - \phi_{\pi}^T(x_3) \right) +2 r_{\pi}
\phi_{\pi}^P(x_2)
 \right]  h_{e}(x_1,x_3,b_1,b_3)
\nonumber\\
& &\;\;\;\;\;\; + 2 r_{\pi}
 \phi_2^P(x_3)a(t')
E_{e}(t') h_{e}(x_3,x_1,b_3,b_1) \bigg\}\;,\label{me6}
\end{eqnarray}
with the factorization scales $t=
\max{\{\sqrt{x_3}m_B,1/b_1,1/b_3\}}$ and  $t'=
\max{\{\sqrt{x_1}m_B,1/b_1,1/b_3\}}$, $r_{K^*_0} =m_{K^*_0} /m_B$.
 The Sudakov form factors $S_B(t)$ and $S_\pi(t)$  and  the   hard
functions $h_{e}$ and others like $h_a$,
 $h_{na}$  are
given explicitly in ref.\cite{kls,lucd}.

Comparing $F^L_{B\to\pi}$ and $F^R_{B\to\pi}$, we find that the
first one is proportional to the small vector decay constant, but
the latter is proportional to the scalar decay constant which is
strongly chiral enhanced, so $F^R_{B\to\pi}$ will give the
dominant contribution.

In ref.~\cite{cheng}, the author found that the vertex corrections
and hard-spectator-scattering corrections can enhance $a_4$
sizably. In PQCD approach, the vertex corrections are at the
next-to-leading order in $\alpha_s$, so we neglect it in our
calculation, but we include the hard spectator scattering (the
last two diagrams in the first row of Fig.~\ref{fig1}). After the
calculation, the non-factorization decay amplitudes for the
$(V-A)(V-A)$ kind of operators read:
\begin{eqnarray}
{\cal M}^L_{B\to\pi}(a) &=& \frac{128 \pi}{ 3\sqrt{6}} m_B^4
\int_0^1 dx_1dx_2dx_3 \int_0^{\infty} b_1 db_1\, b_2
db_2\,\phi_B(x_1,b_1) \phi_{\ks}(x_2)
\nonumber  \\
&&\times \bigg\{-\Big[(x_2-1)\phi_{\pi}(x_3)+r_{\pi}x_3\left(\phi_{\pi}^P(x_3)-\phi_{\pi}^T(x_3)\right)\Big]
a(t)E'_{e}(t) h_n(x_1,1-x_2,x_3,b_1,b_2)
\nonumber \\
&&-\Big[(x_2+x_3)\phi_{\pi}(x_3) -r_{\pi} x_3 \left(
\phi_{\pi}^P(x_3)+\phi_{\pi}^T(x_3)
       \right)\Big]
a(t')E'_{e}(t') h_n(x_1,x_2,x_3,b_1,b_2) \bigg\}.\label{me3}
\end{eqnarray}
For the $(V-A)(V+A)$ kind of operators, the decay amplitudes for
these two diagrams are given by:
\begin{eqnarray}
{\cal M}^R_{B\to\pi}(a) &=& \frac{128 \pi}{ 3\sqrt{6}}m_B^4r_{\ks}
\int_0^1 dx_1dx_2dx_3 \int_0^{\infty} b_1 db_1\, b_2
db_2\,\phi_B(x_1,b_1)\bigg\{a(t) E'_{e}(t)\times
\nonumber \\
& &  \left[ (x_2-1)\left(\phi_{\pi}^A(x_3) ( \phi_{\ks}^S(x_2) +
\phi_{\ks}^T(x_2)) \right.+r_{\pi}(
\phi_{\pi}^P(x_3)-\phi_{\pi}^T(x_3))( \phi_{\ks}^S(x_2)+
\phi_{\ks}^T(x_2))\right)\nonumber\\
& &\;\;\;\;\; \left. -r_{\pi} x_3 \left(
\phi_{\pi}^P(x_3)+\phi_{\pi}^T(x_3) \right) \left(
\phi_{\ks}^S(x_2)- \phi_{\ks}^T(x_2) \right)\right]
h_n(x_1,1-x_2,x_3,b_1,b_2) +\nonumber \\
& &  \left[ x_2\phi_{\pi}^A(x_3) \left( \phi_{\ks}^S(x_2) -
\phi_{\ks}^T(x_2) \right) \right.+r_{\pi}x_2\left(
\phi_{\pi}^P(x_3)-\phi_{\pi}^T(x_3) \right) \left(
\phi_{\ks}^S(x_2)- \phi_{\ks}^T(x_2)
\right)\nonumber\\
& &  \left. + r_{\pi} x_3 \left( \phi_{\pi}^P(x_3)+\phi_{\pi}^T(x_3)
\right) \left( \phi_{\ks}^S(x_2)+ \phi_{\ks}^T(x_2) \right)\right]
a(t')E'_{e}(t') h_n(x_1,x_2,x_3,b_1,b_2)\bigg\},\label{me5}
\end{eqnarray}
where $E'_{e}(t)=\alpha_s(t) \exp[-S_B(t)-S_{2}(t)-S_{3}(t)] $,
the factorization scales are chosen by
\begin{eqnarray}
t&=&
\max{\{\sqrt{x_1x_3}m_B,\sqrt{|(1-x_1-x_2)x_3|}m_B,1/b_1,1/b_2\}},
\\
t'&=&\max{\{\sqrt{x_1x_3}m_B,\sqrt{|(x_1-x_2)x_3|}m_B,1/b_1,1/b_2\}}.
\end{eqnarray}

From the above formulae, we can see that the two hard spectator
scattering diagrams contribute constructively for ${\cal
M}^L_{B\to\pi}$ while most contributions cancelled  for  ${\cal
M}^R_{B\to\pi}$. So it is expected that the $(V-A)(V-A)$ kind
operator contribution can give an important contribution as in
$B\to f_0K$ \cite{f0K}. However, these diagrams are suppressed
compared with
 the factorizable ones for a smaller Wilson coefficient $C_3$,
thus they cannot play such an important role as in QCDF
\cite{cheng}.

In PQCD approach, the annihilation type diagrams are free of
endpoint singularity, so they can be calculated systematically.
The Feynman diagrams are plotted in the second row of
Fig.~\ref{fig1}. For the first two   factorizable annihilation
diagrams, the decay amplitude formulae are written as:
\begin{eqnarray}
F^L_{a}(a) &=& \frac{32 \pi}{ 3} m_B^4f_B \int_0^1 dx_2 dx_3
\int_0^{\infty} b_2db_2\, b_3db_3\
\nonumber \\
&& \times \Bigg\{\Big[(x_3-1)\phi_{\pi}^A(x_3)\phi_{\ks}(x_2) -
2r_{\pi} r_{\ks}(x_3-2) \phi_{\pi}^P(x_3)\phi_{\ks}^S(x_2)
\nonumber \\
&&~~+2r_{\pi} r_{\ks}x_3\phi_{\ks}^S(x_2) \phi_{\pi}^T(x_3)\Big]
a(t) E_{a}(t) h_{a}(x_2,1-x_3, b_2, b_3)
\nonumber\\
&&
 + \Big[x_2\phi_{\pi}^A(x_3)\phi_{\ks}(x_2)-2r_{\pi}
r_{\ks}\phi_{\pi}^P(x_3)((x_2+1)\phi_{\ks}^S(x_2)+(x_2-1)\phi_{\ks}^T(x_2))\Big]\nonumber\\
&& \,\,\,\times a(t') E_{a}(t') h_{a}(1-x_3,x_2, b_3,
b_2)\Bigg\},\label{ma24}
\end{eqnarray}
for the $(V-A)(V-A)$ kind of operators and
\begin{eqnarray}
\nonumber F^R_{a}(a) &=& -\frac{64 \pi}{ 3} m_B^4 f_B \int_0^1
dx_2 dx_3 \int_0^{\infty} b_2db_2\, b_3db_3\,  \Bigg\{E_{a}(t)
a(t)h_{a}(x_2,1-x_3, b_2,
b_3)  \\
\nonumber &&
\times\Big[r_{\pi}(x_3-1)\phi_{\ks}(x_2)\left(\phi_{\pi}^P(x_3)+\phi_{\pi}^T(x_3)\right)
+2r_{\ks}\phi_{\pi}(x_3)\phi_{\ks}^S(x_2)\Big] \\ && -
\Big[2r_{\pi}\phi_{\ks}(x_2)\phi_{\pi}^P(x_3)+r_{\ks}x_2\phi_{\pi}(x_3)
\left(\phi_{\ks}^T(x_2)-\phi_{\ks}^S(x_2)\right)\Big]\nonumber \\
&& \times a(t')E_{a}(t') h_{a}(1-x_3,x_2, b_3, b_2)
\Bigg\},\label{ma6}
\end{eqnarray}
for the $(V-A)(V+A)$ kind of operators, where
\begin{eqnarray}
E_{a}(t)&=&\alpha_s(t) \exp[-S_2(t)-S_{3}(t)],
\end{eqnarray}
with $t= \max{\{\sqrt{1-x_3}m_B,1/b_2,1/b_3\}}$ and  $t'=
\max{\{\sqrt{x_2}m_B,1/b_2,1/b_3\}}$.

For the non-factorizable annihilation diagrams, e.g., the last two
diagrams in the second row of Fig. 1, the factorization formulae
read:
\begin{eqnarray}
 {\cal M}^L_{a}(a) &=& \frac{128 \pi}{ 3\sqrt{6}}m_B^4 \int_0^1
dx_1dx_2dx_3 \int_0^{\infty} b_1 db_1\, b_2 db_2\,\phi_B(x_1,b_1)
 \bigg\{ \left[ -x_2\phi_{\pi}^A(x_3) \phi_{\ks}(x_2)
\right.\nonumber \\
& &\;\;\;\;\;\;+r_{\pi}r_{\ks} \phi_{\pi}^P(x_3) \left(
(x_2-x_3+3)\phi_{\ks}^S(x_2)+(x_2+x_3-1) \phi_{\ks}^T(x_2)
\right)\nonumber\\
& &\;\;\;\;\;\;\left. + r_{\pi}r_{\ks} \phi_{\pi}^T(x_3) \left(
(1-x_2-x_3)\phi_{\ks}^S(x_2)+(1-x_2+x_3) \phi_{\ks}^T(x_2)
\right)\nonumber\right] \\
 & & \,\,\,\,\,\,\,\;\;\times a(t)E'_{e}(t)
h_{na}(x_1,x_2,x_3,b_1,b_2) - \left[ (x_3-1)\phi_{\pi}^A(x_3)
\phi_{\ks}(x_2)
\right.\nonumber \\
& &\;\;\;\;\;\;+r_{\pi}r_{\ks} \phi_{\pi}^P(x_3) \left(
(x_2-x_3+1)\phi_{\ks}^S(x_2)-(x_2+x_3-1) \phi_{\ks}^T(x_2)
\right)\nonumber\\
& &\;\;\;\;\;\;\left. + r_{\pi}r_{\ks} \phi_{\pi}^T(x_3) \left(
(x_2+x_3-1)\phi_{\ks}^S(x_2)-(1+x_2-x_3) \phi_{\ks}^T(x_2)
\right)\nonumber\right] \\ & & \,\,\,\,\,\,\,\,\,\times
a(t')E'_{e}(t') h'_{na}(x_1,x_2,x_3,b_1,b_2)
\bigg\}\;,\label{ma13}
\end{eqnarray}
and for the $(V-A)(V-A)$ kind of operators
 \begin{eqnarray}
 {\cal M}^R_{a}(a)&=& \frac{128 \pi}{ 3\sqrt{6}}m_B^4 \int_0^1
dx_1dx_2dx_3 \int_0^{\infty} b_1 db_1\, b_2
db_2\,\phi_B(x_1,b_1)\times
\nonumber \\
&&\bigg\{ \left[r_{\pi} (x_3-1)\phi_{\ks}(x_2)
(\phi_{\pi}^T(x_3)-\phi_{\pi}^P(x_3))
+r_{\ks}x_2\phi_{\pi}(x_3)(\phi_{\ks}^S(x_2)+\phi_{\ks}^T(x_2))\right]\nonumber \\
&& \;\;\times a(t)E'_{e}(t) h_{na}(x_1,x_2,x_3,b_1,b_2)
\nonumber \\
&& - \left[r_{\pi} (x_3-1)\phi_{\ks}(x_2)
(\phi_{\pi}^T(x_3)-\phi_{\pi}^P(x_3))+r_{\ks}x_2\phi_{\pi}(x_3)(\phi_{\ks}^S(x_2)
+\phi_{\ks}^T(x_2))
\right]\nonumber \\
&&\;\;\times a(t')E'_{e}(t') h'_{na}(x_1,x_2,x_3,b_1,b_2)
\bigg\}\;,\label{ma5}
\end{eqnarray}
for the $(V-A)(V+A)$ kind of operators, where
\begin{eqnarray}
   t &=& \nonumber
\max{\{\sqrt{x_2(1-x_3)}m_B,\sqrt{1-(1-x_1-x_2)x_3}m_B,1/b_1,1/b_2\}},\\
\nonumber t'&=&
\max{\{\sqrt{x_2(1-x_3)m_B},\sqrt{|(x_1-x_2)(1-x_3)|}m_B,1/b_1,1/b_2\}}.
\end{eqnarray}

Summing up all contributions mentioned above, the decay amplitude
for $B \to \bk0\pi^- $ is
 \begin{eqnarray}
 A(B^-\to
\bk0\pi^-)\nonumber &&
=\frac{G_F}{\sqrt{2}}\bigg[V_{ub}V_{us}^*\bigg\{F^L_{a} ( a_1)
+{\cal M}^L_{a} (C_1)\bigg\}\nonumber\\
&&-V_{tb}V_{ts}^*\bigg\{F^L_{e} (a_4-\frac{1}{2}a_{10})
+F^R_{e}(a_6-\frac{1}{2}a_8)+F^L_{a} (a_4+a_{10})+F^R_{a} (a_6+a_8)\nonumber \\
&&\;\;+ {\cal M}^L_{e } (C_3-\frac{1}{2}C_9)+{\cal M}^R_{e}
(C_5-\frac{1}{2}C_7)+{\cal M}^L_{a} (C_3-\frac{1}{2}C_9)+{\cal
M}^R_{a} (C_5-\frac{1}{2}C_7) \bigg\}\bigg]\label{kpp},
\end{eqnarray}
where the combinations of Wilson coefficients are defined as usual
\cite{akl}:
\begin{eqnarray}
a_1= C_2+C_1/3, & a_3= C_3+C_4/3,~a_5= C_5+C_6/3,~a_7=
C_7+C_8/3,~a_9= C_9+C_{10}/3,\\
a_2= C_1+C_2/3, & a_4= C_4+C_3/3,~a_6= C_6+C_5/3,~a_8=
C_8+C_7/3,~a_{10}= C_{10}+C_{9}/3.
\end{eqnarray}
 For the decays $\bar B^0\to
K_0^{*-}\pi^+$, $B^-\to K_0^{*-} \pi^0$ and $\bar B^0 \to
\bk0\pi^0$, the analysis is similar, except the last two channels
include the $\pi$-emission diagrams in the third row of
Fig.~\ref{fig1}. The decay amplitudes of the factorizable
$\pi$-emission diagrams for the $(V-A)(V-A)$ kind of operators are
written by:
 \bea F^L_{B\to K^*_0}(a) &=&
\frac{32 \pi}{ 3} m_B^4 f_{\pi}\int_0^1 dx_1 dx_2 \int_0^{\infty}
b_1db_1\, b_2db_2\, \phi_B(x_1,b_1)\bigg\{a(t) E_{e}(t)
\nonumber \\
& &\times  \left[ (1+x_2)\phi_{\ks}(x_2)-r_{\ks}(1-2x_2) \left(
\phi_{\ks}^s(x_2)+\phi_{\ks}^T(x_2) \right) \right]
h_{e}(x_1,x_2,b_1,b_2)
\nonumber\\
& &\;\;\;\;\;\; -2r_{\ks} \phi_{\ks}^s({x_2})a(t')E_{e}(t')
h_{e}(x_2,x_1,b_2,b_1) \bigg\},\label{me19}
\end{eqnarray} and for the $(V-A)(V+A)$ kind of operators:
 \bea F^R_{B\to K^*_0}(a)=-F^L_{B\to K^*_0}(a).\eea
For the non-factorizable diagrams $(V-A)(V-A)$ operators:
 \begin{eqnarray}
 {\cal M}^L_{B\to K^*_0}(a) &=& \frac{128 \pi}{ 3\sqrt{6}}m_B^4 \int_0^1
dx_1dx_2dx_3 \int_0^{\infty} b_1 db_1\, b_3 db_3\,\phi_B(x_1,b_1)
\phi_{\pi}^A(x_3)\bigg\{a(t)E'_{e}(t)
\nonumber \\
&&
\bigg[(1-x_3)\phi_{\ks}(x_2)+r_{\ks}x_2\bigg(\phi_{\ks}^s(x_2)-\phi_{\ks}^T(x_2)\bigg)\bigg]
 h_n(x_1,1-x_3,x_2,b_1,b_3)
\nonumber \\
&&+\left[-(x_2+x_3)\phi_{\ks}(x_2) -r_{\ks} x_2
(\phi_{\ks}^s(x_2)+\phi_{\ks}^T(x_2))
       \right]\nonumber
        \\
&& a(t') E'_{e}(t') h_n(x_1,x_3,x_2,b_1,b_3) \bigg\}\;,\label{me24}
 \end{eqnarray}
and for $(V-A)(V+A)$ operators
\begin{eqnarray}
 {\cal M}^R_{B\to K^*_0}(a) &=& \frac{128 \pi}{ 3\sqrt{6}}m_B^4 \int_0^1
dx_1dx_2dx_3 \int_0^{\infty} b_1 db_1\, b_3 db_3\,\phi_B(x_1,b_1)
\phi_{\pi}^A(x_3)\bigg\{-a(t)E'_{e}(t)
\nonumber \\
&&\bigg[(x_2-x_3+1)\phi_{\ks}(x_2)+r_{\ks}x_2(\phi_{\ks}^s(x_2)+\phi_{\ks}^T(x_2))\bigg]
 h_n(x_1,1-x_3,x_2,b_1,b_3)
\nonumber \\
&&+\left[x_3\phi_{\ks}(x_2) +r_{\ks} x_2
(\phi_{\ks}^s(x_2)-\phi_{\ks}^T(x_2))
       \right]
a(t')E'_{e}(t') h_n(x_1,x_3,x_2,b_1,b_3) \bigg\}\;.\label{me8}
\end{eqnarray}
 We
write the decay amplitudes for the other three channels below:
\begin{eqnarray}
 \nonumber  A(\bar B^0\to
K_0^{*-}\pi^+)&&=\frac{G_F}{\sqrt{2}}\Big[V_{ub}V_{us}^*\left\{F^L_{B\to\pi}(a_1)+{\cal
M}^L_{B\to\pi}(C_1)\right\}\nonumber\\
&&-V_{tb}V_{ts}^*\Big\{F^L_{B\to\pi}(a_4+a_{10})+F^R_{B\to\pi}(a_6+a_8
)+F^L_{a} (a_4-\frac{1}{2}a_{10} )+F^R_{a} (a_6-\frac{1}{2}a_8
)\nonumber\\&&\;\;\;\;+{\cal M}^L_{B\to\pi}(C_3+C_9 )+{\cal
M}^R_{B\to\pi} (C_5+C_7 )+{\cal M}^L_{a} (C_3-\frac{1}{2}C_9
)+{\cal M}^R_{a}(C_5-\frac{1}{2}C_7)\Big\}\Big],\label{k0p}
\\
A(B^-\to K_0^{*-} \pi^0) \nonumber
&&=\frac{G_F}{{2}}\Big[V_{ub}V_{us}^*\left\{F^L_{B\to\pi}(a_1
)+{\cal M}^L_{B\to\pi}(C_1)+F^L_{B\to K^*_0}(a_2)+{\cal M}^L_{B\to
K^*_0} (C_2 )+F^L_{a}(a_1)+{\cal
M}^L_{a} (C_1 )\right\}\\
&&-V_{tb}V_{ts}^*\Big\{F^L_{B\to\pi}(a_4+a_{10})+F^R_{B\to\pi}(a_6+a_8
)+F^L_{B\to K^*_0}(\frac{3}{2}a_9-\frac{3}{2}a_7)\nonumber\\
&&\;\;\; +F^L_{a} (a_4+a_{10}
)+F^R_{a} (a_6+a_8 )\nonumber \\
&&\;\;\;+{\cal M}^L_{B\to\pi}(C_3+C_9 )+{\cal M}^R_{B\to\pi}
(C_5+C_7 )+{\cal
M}^L_{B\to K^*_0}(\frac{3}{2}C_{10} )+{\cal M}^R_{B\to K^*_0}(\frac{3}{2}C_8 )\nonumber\\
&&\;\;\;+{\cal M}^L_{a} (C_3+C_{10} )+{\cal
M}^R_{a}(C_5+C_7)\Big\}\Big],\label{kp0}
\\
A(\bar B^0 \to \bk0\pi^0) \nonumber
&&=\frac{G_F}{{2}}\Big[V_{ub}V_{us}^*\left\{F^L_{B\to
K^*_0}(a_2)+{\cal
M}^L_{B\to K^*_0} (C_2 )\right\}\nonumber\\
&&
\;\;\;-V_{tb}V_{ts}^*\Big\{-F^L_{B\to\pi}(a_4-\frac{1}{2}a_{10})
-F^R_{B\to\pi}(a_6-\frac{1}{2}a_8 )+F^L_{B\to K^*_0}(\frac{3}{2}a_9)+F^R_{B\to K^*_0}(\frac{3}{2}a_7) \\
&& \;\;\;\;-F^L_{a} (a_4-\frac{1}{2}a_{10} )-F^R_{a} (a_6-\frac{1}{2}a_8 ) \nonumber\\
&&\;\;\;-{\cal M}^L_{B\to\pi}(C_3-\frac{1}{2}C_9 )-{\cal
M}^R_{B\to\pi}(C_5-\frac{1}{2}C_7 )+{\cal M}^L_{B\to
K^*_0}(\frac{3}{2}C_{10} )
 +{\cal M}^R_{B\to K^*_0}(\frac{3}{2}C_8)\nonumber\\
&& \;\;\;-{\cal M}^L_{a} (C_3-\frac{1}{2}C_9 ) -{\cal
M}^R_{a}(C_5-\frac{1}{2}C_7) \Big\}\Big].\label{k0p0}
\end{eqnarray}
The isospin relation for the four channels holds exactly in these
equations:
\begin{eqnarray}
\sqrt{2} A(\bar B^0 \to \bk0\pi^0) +  A(\bar B^0\to
K_0^{*-}\pi^+)=\sqrt{2}A(B^-\to K_0^{*-} \pi^0)-A(B^-\to
\bk0\pi^-).
\end{eqnarray}

\subsection{$B \to a_0K$ decays}

As mentioned above, the predicted branching fractions of $B \to
a_0K$, which overshoot the experimental limits, are regarded as an
evidence to rule out scenario I in QCDF, thus it is important to
see whether it is the same in the PQCD approach.

The Feynman diagrams for these decays are completely the same as
the $B\to\ks \pi$ except that we should identify the $s\bar{q}$
and $q\bar{q}$ as $K$ and $a_0$ rather than $\ks$ and $\pi$, then
each channel corresponds to the one in $B\to\ks \pi$. Their
factorization formulae can be derived from the corresponding
channels directly. The $K$-emission and annihilation decay
amplitudes of $B \to a_0K$ can be obtained from $B \to \ks\pi $ by
making the substitution: \bea &&\nonumber \phi_{\ks}\to
\phi^A_{K},\phi^S_{\ks}\to
\phi^P_{K},\phi^T_{\ks}\to \phi^T_K, \\
 &&\phi^A_{\pi}\to
\phi_{a_0},\phi^P_{\pi}\to- \phi^S_{a_0},\phi^T_{\pi}\to
-\phi^T_{a_0}.
\end{eqnarray}
The $a_0$-emission diagrams have only nonfactorizable
contributions, the substitution for $(V-A)(V-A)$ operators is:
 \bea
 &&\phi^A_{\pi}\to
\phi_{a_0},\phi^P_{\pi}\to \phi^S_{a_0},\phi^T_{\pi}\to
\phi^T_{a_0},\\
&&\nonumber \phi_{\ks}\to \phi^A_{K},\phi^S_{\ks}\to-
\phi^P_{K},\phi^T_{\ks}\to -\phi^T_K,  \\
\end{eqnarray} but for $(V-A)(V+A)$ operators,
 \bea
 &&\phi^A_{\pi}\to
\phi_{a_0},\phi^P_{\pi}\to \phi^S_{a_0},\phi^T_{\pi}\to
\phi^T_{a_0},\\
&&\nonumber \phi_{\ks}\to -\phi^A_{K},\phi^S_{\ks}\to
\phi^P_{K},\phi^T_{\ks}\to \phi^T_K.  \\
\end{eqnarray}

Compared with $B\to\ks \pi$, the features of  $B\to a_0K$ are:
\begin{itemize}
\item For the decays $\bar{B^0}\to a_0^+ K^-$ and $B^-\to a_0^-
\bar{K}^0$, the emitted particle is $K$, which can be produced
through the axial-vector current without any suppression, thus the
operator $O_4$ can give a large contribution to the emission
factorizable amplitudes. But this term has a minus sign relative
to $O_6$, so the penguin operators cancel with each other sizably.
The contribution from tree operators can be large due to the large
Wilson coefficients $a_1$ in $B^-\to a_0^+K^-$.

\item In $B\to \bk0\pi^-$, the emitted particle ($\bk0$) is a
scalar meson, the two hard spectator scattering diagrams
(non-factorizable) can enhance each other due to the
anti-symmetric twist-2 distribution amplitudes. But in
$\bar{B^0}\to a_0^+ K^-$ and $B^-\to a_0^- \bar{K}^0$, the emitted
particle is a pseudoscalar, there are cancellations between the
two hard spectator scattering diagrams. So the hard spectator
scattering contribution is rather small.

\item The annihilation diagrams of the four $B\to a_0K$ channels
are similar with each other, the dominant contributions are all
from the $S\to P$ time-like form factor mediated by an $(S+P)$
density.

\item $\bar{B^0}\to a_0^0 \bar K^0$ and $B^-\to a_0^0 K^-$ are
more complicated due to the appearance of the $a_0$-emission
diagrams. Because of  the vanishing vector decay constant of
$a_0$, the factorizable emission diagrams are zero. For the
nonfactorizable diagrams, the QCD penguin operators cancel for the
neutral state of isospin triplet. The electroweak penguin
operators have a small Wilson coefficients, thus the emission
contributions in these channels are  small. For the tree
operators, although they are suppressed by the CKM matrix
elements, the nonfactorizable emission diagrams can be enhanced
for the large Wilson coefficients $C_2$. So it is expected a large
CP asymmetry in the decays $\bar{B^0} \to a^0_0 \bar K^0$ and $B^-
\to a^0_0 K^-$.
\end{itemize}
From the above discussion, we can see that the dominant
contributions are from the annihilation diagrams and the diagrams
with tree operators.


\section{Results and discussions}
\label{sec4}
\begin{table}\caption{Input parameters used in the numerical calculation}
\begin{center}
\begin{tabular}{c|ccc}
\hline \hline
 Masses &$m_{\ks}=1.412 \mbox{ GeV}$,   &$ m_0^K=1.7 \mbox{ GeV}$, &$m_0^\pi=1.4 \mbox{ GeV}$ \\
 & $m_{a_0}=0.98{\mbox{ GeV}}$  & $ M_B = 5.28 \mbox{ GeV}$&\\
 \hline
  Decay constants &$f_B = 0.19 \mbox{ GeV}$  & $ f_{K} = 0.16
 \mbox{ GeV}$ & $f_\pi=0.132\mbox{ GeV}$\\
 \hline
Life Times &$\tau_{B^\pm}=1.671\times 10^{-12}\mbox{ s}$ &
$\tau_{B^0}=1.536\times 10^{-12}\mbox{ s}$&\\
 \hline
CKM &$V_{tb}=0.9997$ & $V_{ts}=-0.04$,&\\
 &$V_{us}=0.2196$ & $V_{ub}=0.00367e^{-i60^{\circ}}$&\\
\hline \hline
\end{tabular}\label{para}
\end{center}
\end{table}

For numerical calculations, we have employed the parameters in
Tab.~\ref{para}. For the $B$ meson wave function, we adopt the
Gaussian-type model \cite{kls} (we choose the shape parameter
$\omega=0.4$). As for the light-cone distribution amplitudes
(LCDAs) of the pion and kaon, we use the results from QCD sum
rules up to twist-3 \cite{QCDSR}. Other parameters relevant to the
scalar mesons have been given in the second section.

\subsection{The Branching Ratios and The CP Asymmetries}

\begin{table}\caption{ Various decay amplitudes ($\times 10^{-2}\mbox{GeV}^3$) in decay $B^-\to \bk0
\pi^-$ and $B^-\to a_0^- \bar{K}^0$}
\begin{tabular}{c|c|c|c|c|c}\hline \hline & $F^L_{e}(a_4)$ & $F^R_{e} (a_6)$
 & ${\cal M}^L_{e}(C_3)+M_e^R(C_5)$
 &$F^L_a(a_4)+F_a^R(a_6)+{\cal
M}_a^L(C_3)+{\cal M}_a^R( C_5)$ & $F^L_{a}(a_2)+{\cal M}^L_{a}(C_1) $\\
\hline
  scenario I &$0.98 $&$-12$&$0.78 +1.2i$
  &$7.4+13.3 i$
  &$8.8 -11.8i$\\
   scenario II &$-1.4$& $17.9$&$2.1-0.37 i$
  &$-5.8 -17.8 i$&$-13.6-0.11 i$\\ \hline
  $B^-\to a_0^- \bar{K}^0$ &$9.3 $&$-11.3 $&$0.09 -0.74
  i$ &$2.0-9.0i$
  &$8.6+1.2i$  \\ \hline\hline
\end{tabular}\label{t1}
\end{table}

 Using the parameters in the above, we give the numerical results
for different amplitudes of $B^- \to\bar K^{*0}_0 \pi^-$ in Table
\ref{t1}. The numerical results confirm that the emission diagram
of $(V-A)(V-A)$ operators $F_{B\to\pi}^L(a_4)$ indeed give small
contributions because of the small vector current decay constant.
The $(V-A)(V+A)$ operators give the dominant contribution
$F_e^R(a_6)$. The opposite sign between scenario I and scenario II
comes from the decay constant of the $\ks$ meson. The
non-factorizable contribution ${\cal M}^L_{B\to\pi}(C_3)$ is small
due to the small Wilson coefficient. ${\cal M}^R_{B\to\pi}(C_5)$
is even smaller because of the cancellation between the two
diagrams.

According to PQCD power counting, the annihilation diagrams are
power suppressed, but the suppression is not so effective in some
cases, such as when chiral enhancement existing. Usually there is
a large imaginary part in the amplitudes of the annihilation
diagrams, which is the source of strong phase in PQCD approach.
The numerical results in Table \ref{t1} indicate that the
annihilation diagrams in scenario I are more important than that
in scenario II. There are also tree operators contributing to the
annihilation diagrams $F^L_{a}(a_2)+{\cal M}^L_{a}(C_1)$, which
are Cabibbo suppressed, but they are essential in direct $CP$
violation.


\begin{table}\caption{Branching ratios for the decays $B\to \ks \pi$ and $B \to a_0 K$(in units of
$10^{-6}$). The first theoretical error is from the decay constant
of the scalar meson, the second and the third  one is Gengebauer
moments $B_1$ and $B_3$, the uncertainty caused the CKM angle
$\gamma$ is very small which is not listed here. The experimental
data listed here are the world average values by the Heavy Flavor
Averaging Group (HFAG) \cite{hfag}.}
\begin{tabular}{c|c|c|c||c|c|c}
\hline \hline
 Channel &   {scenario I}&  {scenario
  II} &  {exp.}& Channel &   {scenario
  I} & \multicolumn{1}{c}{exp.}\\
 \hline
  $B^- \to \bk0 \pi^-$ & $20.7^{+4.3+0.8+1.8}_{-3.9-0.8-1.6}$&
    $47.6^{+11.3+3.7+6.9}_{-10.1-3.6-5.1}$& $41.2\pm 4.2$&
    $B^- \to \bar{K}^0 a_0^-$  &$6.9^{+0.8+1.1+2.0}_{-0.7-1.1-1.7}$& $<3.9$  \\
$\bar B^0 \to \km \pi^+$  &$20.0^{+4.2+0.8+1.6}_{-3.8-0.7-1.5}$&
    $43.0 ^{+10.2+3.1+7.0}_{-9.1-2.9-5.2} $&$46.6^{+5.6}_{-6.6}$&
    $\bar B^0 \to K^-a_0^+$  &$9.7^{+1.1+1.6+2.7}_{-1.0-1.4-2.2}$&$<1.6$\\
  $\bar B^0 \to \bk0 \pi^0$ & $10.0^{+2.1+0.4+1.0}_{-1.9-0.5-0.9}$
  &$18.4^{+4.4+1.5+4.0}_{-3.9-1.4-2.9} $&$25.5\pm9.9$ &$\bar B^0 \to
\bar{K}^0 a_0^0$&$4.7^{+0.5+0.7+1.1}_{-0.5-0.8-1.1}$&$<7.8$\\
  $B^- \to \km \pi^0$  &$11.3^{+2.4+0.4+0.7}_{-2.1-0.3-0.7}$&
  $28.8^{+6.8+1.9+3.2}_{-6.1-1.9-3.5}$&-&$B^- \to K^- a_0^0$
  &$3.5^{+0.4+0.4+1.0}_{-0.4-0.6-1.0}$&$<2.5$\\
\hline\hline
\end{tabular}\label{t3}
\end{table}

Now it is straightforward to obtain the results for the
$CP$-averaged branching ratio of $B^- \to\bar K^{*0}_0 \pi^-$,
which is given in Tab. \ref{t3}. Comparing the two scenarios, we
find: the results from scenario II are twice as scenario I, the
most important reason is the larger scalar decay constant in
scenario II; secondly, the nonfactorizable diagrams are small,
they only change the branching ratio slightly; the annihilation
diagrams play an important role in both scenarios, it can enhance
the branching ratios about $50\%$ in scenario I, and about $30\%$
in scenario II. The current experimental data \cite{hfag} is also
listed in Tab.~\ref{t3}. The large branching ratio is consistent
with the results in scenario II, so scenario II is more preferable
than scenario I, this conclusion is consistent with \cite{cheng}.
But the difference with \cite{cheng} is: we directly calculate the
annihilation contribution in $B\to \ks\pi$, rather than fit the
$B\to \ks\pi$ data and then use the $B\to a_0K$ data to rule out
scenario I.

The decay amplitudes for the $B^- \to a_0^-\bar{K^0}$ decay are
also listed in table \ref{t1}, the results indicate that the
emission diagrams almost cancelled out, as expected in section
III. The branching ratio is dominated by the annihilation diagrams
which is at the same level as the $B^- \to \bk0 \pi^-$, and the
induced branching ratio is about twice larger than the
experimental upper bound. The results also show that scenario I is
not supported by the current experimental data.

In the above discussion, we concentrate on $B^- \to \bk0 \pi^-$
and $B \to a_0^-\bar{K^0}$. The dominant contribution in $B\to \ks
\pi$ for other channels is the same with $B^- \to \bk0 \pi^-$ by isospin symmetry,
 except the contribution from the electro-weak penguin
and the tree operators which can violate isospin symmetry. To
explore the deviation of the isospin limit, it is convenient to
define the parameters below: \bea \nonumber &&R_1=\frac{{\cal
B}({\bar{B}^0\to \k0 \pi^0})}{{\cal B}({\bar{B}^0\to \km \pi^+})},\\
\nonumber && R_2=\frac{{\cal B}({B^-\to \km\pi^0})}{{\cal
B}({B^-\to \k0 \pi^-})},\\
&& R_3=\frac{\tau(B^0)}{\tau(B^-)}\frac{{\cal B}({B^-\to
\k0\pi^-})}{{\cal B}({\bar{B}^0\to \km\pi^+})}.
\end{eqnarray}
For $B \to a_0K$, the definition is the similar except $\ks\to K,
\pi \to a_0$. These parameters are the ratios of the branching
fractions, which should be less sensitive to many nonperturbative
inputs than the branching fractions, thus it is more persuadable
to test these parameters. In the isospin limit, if we ignore the
$CKM$ suppressed tree diagrams and electro-weak penguins, $R_1$,
$R_2$ and $R_3$ should be equal to 0.5, 0.5 and 1.0.  The
deviations reflect the magnitude of the tree operators and the
electro-weak penguins directly. Our results and the experimental
data are given in table \ref{t4}, where we use the central values
in table~\ref{t3} for the experiment data. In both scenarios, the
deviations from isospin limit are not large, which shows that the
QCD penguin are dominant in the branching ratios, both in emission
diagrams and the annihilation diagrams. The three ratios for $B
\to a_0 K$ decays are: $R_1=0.48, R_2=0.51, R_3=0.68$. There is a
large deviation for the ratio $R_3$, and the reason is the large
tree contribution. So the large direct CP asymmetry for $B^0\to
a^-_0K^+$ is also expected.

\begin{table}\caption{Ratios of the branching fractions in $B \to \ks
\pi$. For the experimental values, we only use the central
values.}
\begin{tabular}{c|c|c|c|c}\hline \hline  Ratios&isospin limit &
 Scenario II & Scenario I & Experiment\\
\hline
 $R_1$&0.5&0.43&0.48&0.39 \\
 $R_2$&0.5&0.61&0.55& -\\
 $R_3$&1.0&1.02&0.95&0.81 \\ \hline\hline
  \end{tabular}\label{t4}
\end{table}

The contribution from different effective operators shown in
Eqs.~(\ref{kpp}-\ref{k0p0}) have been categorized to two groups
according to the different $CKM$ matrix elements:
 \bea
\bar{A}=\frac{G_F}{\sqrt{2}}[V_{ub}V_{us}^*T-V_{tb}V_{ts}^*P],
\label{amp}
 \eea
where $T/P$ denotes the amplitude which comes from the
tree/penguin operators respectively. The charge conjugate channel
decay amplitude is the same as Eq.~(\ref{amp}) except the sign of
the weak phase.
 The formula for the direct CP asymmetry reads:
 \bea
 A_{CP}=\frac{|\bar{A}|^2-|A|^2}{|\bar{A}|^2-|A|^2}
=\frac{2z\sin{\gamma}\sin{\delta}}{1-2z\cos{\gamma}\cos{\delta}+z^2},
\eea where
$z=|\frac{V_{tb}V_{ts}^*}{V_{ub}V_{us}^*}||\frac{P}{T}|$, $\delta$
is the relative strong phase between two groups of contributions.
This equation indicates that the direct $CP$ violation depends on
the ratio of the tree and penguin contribution. The direct $CP$
asymmetry is very small if the ratio is too large or too small,
while the comparable tree and penguin contributions imply large
direct $CP$ asymmetries. For $B \to \ks\pi$, the penguin operators
give the dominant contribution, but the tree operators suffer from
the  $CKM$ suppression, so it is expected the direct $CP$
asymmetry is small. We list our results in table \ref{t6} as well
as the experimental data \cite{hfag}, where the results are
consistent with the experiments in both scenarios. But in the
decay $ B^0 \to a_0K$, as mentioned above, the emission penguin
contribution cancels, while the tree operators are large, sizable
direct $CP$ asymmetries are predicted in these channels,
especially in $\bar{B^0}\to a_0^+K^-$ and $B^-\to a_0^0K^-$.
 In QCDF, the
central values of direct $CP$ asymmetries for all four channels
are very small, but with large uncertainties for $B^-\to a_0^0K^-$
and $\bar B^0\to a_0^+K^-$. Furthermore, we may expect the similar
size of $CP$ asymmetries in similar decays $B\to a_0(1450)K$.

\begin{table}\caption{Direct CP asymmetries (in units of \%) }\label{t6}
\begin{tabular}{c|c|c|c|c|c}
\hline \hline
Channel & Scenario I & Scenario II & exp.&channels& $A_{CP}$ \\
\hline   $B^- \to \bk0 \pi^-$ &$-1.5$&$-1.7$ &
 $-5^{+5}_{-8}$&$B^- \to \bar{K}^0 a_0^-$& 4\\
   $\bar B^0 \to \km \pi^+$ &$9.2$& $0.22$
   & $-7 \pm 14$&$\bar B^0 \to K^-a_0^+$ &-70\\
  $\bar B^0 \to \bk0 \pi^0$&$-9.0$&$-6.8$& $-34\pm 19$& $\bar B^0 \to \bar{K}^0 a_0^0$&-17\\
  $B^- \to \km \pi^0$ &$16.0$& $3.5$& -&$B^- \to K^- a_0^0$&-70\\
\hline\hline
\end{tabular}
\end{table}

\subsection{The Theoretical Uncertainties}

In the above calculation, the uncertainties from the decay
constants can give sizable effects on the branching ratio, but not
to the direct $CP$ asymmetries. Furthermore, there are other
sources of uncertainties:

\begin{itemize}
\item  The twist-3 distribution amplitudes of the scalar mesons
are taken as the asymptotic form for lack of more reasonable
results from non-perturbative methods. These distribution
amplitudes will be studied in the future work \cite{LWZ}. In
\cite{LWZ}, we find the Gegenbauer moments of the twist-3
distribution amplitudes are rather small, which implies the
results will not be changed sizably.

\item The Gegenbauer moments $B_1$ and $B_3$ for twist-2 LCDAs of
$\ks$ and $a_0$ have sizable uncertainties, which can lead to the
theoretical errors. We include these uncertainties in the results
and they can give about $20\% \sim 30\%$ uncertainties to the branching
ratio.

\item The uncertainties of the light pseudoscalar meson and $B$
meson wave functions, the factorization scale, {\it et al.} have
been studied extensively in \cite{kurimoto}. The uncertainty from
the factorization scale is within $10\%$. The major source of the
uncertainty comes from the meson distribution amplitudes. The
results can be varied by $(10-30)\%$ by changing the parameter in
the wave functions.

\item The sub-leading order contributions in PQCD approach have
also been neglected in this calculation, but these corrections
have been calculated in refs. \cite{subleadingPQCD} for $ B\to \pi
K, \pi\pi$, {\it etc.} These corrections can change the penguin
dominated processes about $20\%$ of the branching ratio of $B\to
\pi K$. We may expect similar effect in $B\to \ks \pi$.

\item The uncertainties of $CKM$ matrix elements and the $CKM$
phase angle can also affect  the branching ratios and $CP$
asymmetries. In the two kinds of decays considered in this paper,
the decay amplitudes are the functions of the $CKM$ angle
$\gamma$, whose value given in PDG06 is
$\gamma=(63^{+15}_{-12})^{\circ}$ \cite{pdg06}. With the $CKM$
angle $\gamma$ varying at this area, we find that the error area
is very small in $B \to \ks \pi$ decays in both scenarios. For the
$B \to a_0K$ decays, the error area is some larger, but within ten
percent.

\item The long distance re-scattering can also affect the
branching ratios and $CP$ asymmetries. This effect could be
phenomenologically studied in the final-state interactions
\cite{FSI}. We need more data to determine whether it is necessary
to include the re-scattering effect in $B\to SP$ decays.
\end{itemize}


\section{summary}

In this paper, we calculate the decay modes $B \to \ks \pi$ and $B
\to a_0 K$ within perturbative QCD framework. For $B \to \ks \pi$,
we perform our calculation in two scenarios of the scalar meson
spectrum, our calculation indicates that: scenario II is more
consistent with the experimental data than scenario I. We directly
calculate the contribution from annihilation diagrams: it can
enhance the branching ratios about $50\%$ in scenario I, and about
$30\%$ in scenario II. Our predicted branching ratio of $B \to
a_0K$ in scenario I is larger than the experimental upper bound,
which indicates $a_0(980)$ can not be interpreted as $\bar qq$. We
calculated the direct $CP$ asymmetries and the isospin parameters
in these decays, and we find that in $B \to \ks \pi$ (in both
scenarios) the direct $CP$ asymmetries are small, which are
consistent with the present experiments; the deviation from the
isospin limit is also small. There is large $CP$ asymmetries in $B
\to a_0(980)K$ due to the relatively large tree contributions in
scenario I. We expect similar $CP$ asymmetries in $B \to
a_0(1450)K$ .

\section*{Acknowledgement}

This work was partly supported by the National Science Foundation
of China. We thank Y. Li, Y. M. Wang, M. Z. Yang and H. Zou for
helpful discussions. We would like to acknowledge S.F. Tuan for
useful comments.  C.D. L\"u thanks A. Ali and G. Kramer for their
hospitality during his visit at DESY.



\begin{thebibliography}{99}

\bibitem{scenarioI}G.L. Jaffe, Phys. Rev. D{\bf 15}, 267(1977); $ibid$: {\bf15},
281(1977); S.G. Gorishinii, A.L. Kataev and S.A. Larin, Phys.
Lett. B{\bf135},457(1984); N.N. Achasov, Phys. Usp. {\bf41},
1149(1998), Usp. Fiz. Nauk {\bf168}, 1257 (1998),arXiv:
hep-ph/9904223; A.L. Kataev, Phys. Atom. Nucl. {\bf68}, 567
(2005), Yad. Fiz. {\bf 68}, 597(2005), arXiv: hep-ph/0406305; A.
Vijande, A. Valcarce, F. Fernandez and B. Silvestre-Brac, Phys.
Rev. D{\bf 72}, 034025(2005).

\bibitem{scalarb}
V. Elias, A.H. Fariborz, F. Shi, T.G. Steele, Nucl. Phys. A
{\bf633}, 279(1998); E. van Beveren, Eur. Phys. J.C{\bf10},
469(1999); D. Black, A.H. Fariborz and J. Schechter, Phys. Rev
D{\bf 61}, 074001(2000); E. van Beveren, Phys. Lett. B{\bf495},
300(2000); Erratum-ibid. B{\bf509}, 365(2001); F. Kleefeld, E. van
Beveren and M.D. Scadron, Phys. Rev. D{\bf66}, 034007(2002); M.
Ishida and S. Ishida, arXiv: hep-ph/0310062; M.D. Scadron, G.
Rupp, F. Kleefeld and E. van Beveren, Phys. Rev. D{\bf69},
014010(2004); Erratum-ibid.D{\bf69}, 059901(2004); A.M. Fariborz,
Int. J. Mod. Phys. A{\bf 19}, 2095(2004); Int. J. Mod. Phys. A{\bf
19}, 5417(2004); Phys. Rev. D{\bf 74}, 054030(2006); S. Narison,
Phys. Rev. D{\bf73}, 114024(2006); E. van Beveren, D.V. Bugg, F.
Kleefeld and G. Rupp, Phys. Lett. B{\bf641}, 265(2006).

\bibitem{scalarc}P. Minkowski and W. Ochs, Euro. Phys. J. C{\bf 9}, 283(1999).


\bibitem{scalar}S. Spanier and N. A. T\"ornqvist,
``Note on scalar mesons" in Particle Data Group, Journal of
Physics G {\bf33}, 1(2006); S.Godfrey and J. Napolitano, Rev. Mod.
Phys. {\bf 71},1411(1999); F. E. Close and N. A. T\"ornqvist, J.
Phys. G {\bf 28}, R249(2002).


\bibitem{first}Belle Collaboraion, K. Abe {\it et al.}, Phys. Rev.
D{\bf 65}, 092005(2002).

\bibitem{garmash}
Belle Collaboraion, A. Garmash {\it et al.}, Phys. Rev.D{\bf71}, 092003(2005);
Belle Collabaraion, A. Garmash {\it et
al.}, arXiv: hep-ex/0505048.

\bibitem{babar}
Babar Collaboration, B. Aubert {\it et al.}, Phys. Rev. D{\bf 70},
092001(2004); Phys. Rev. D{\bf 70}, 111102 (2004); Phys. Rev.
D{\bf 72}, 072003(2005); arXiv: hep-ex/0408032, hep-ex/0408073.

\bibitem{belle}Belle Collaboration, K. Abe {\it et al.}, arXiv: hep-ex/0509047;
Belle Collaboration, A. Bonder {\it et al.}, arXiv:
hep-ex/0411004.

\bibitem{abe}Belle Collaboration, K. Abe {\it et al.}, hep-ex/0509001.

\bibitem{new}Babar Collaboration, B. Aubert {\it et al.}, Phys. Rev. D{\bf74}, 032003(2006);
hep-ex/0607112.

\bibitem{hfag}Heavy Flavor Averaging Group, P. Chang, {\it et al.}, http://www.slac.stanford.edu/xorg/hfag.

\bibitem{cheng}H.-Y. Cheng, C.-K. Chua, K.-C. Yang,
Phys. Rev. D{\bf 73}, 014017(2006).

\bibitem{zerobin}A.V. Manohar and I.W. Stewart, arXiv:
hep-ph/0607001.

\bibitem{kls}Y.~Y.~Keum, H.-n. Li and A.~I.~Sanda, Phys. Lett. B{\bf 504}, 6(2001);
Phys. Rev. D{\bf 63}, 054008 (2001); C.H. Chen, Y.Y. Keum and
H.-n. Li, Phys. Rev. D{\bf 66}, 054013(2002).

\bibitem{lucd}C.-D. L\"u, K. Ukai, M.-Z. Yang, Phys. Rev. D{\bf 63}, 074009
(2001); C.~D.~L\"u and M.~Z.~Yang, Eur. Phys. J. C{\bf 23},
275(2002).

\bibitem{luanni}C. D. L\"u, K. Ukai, Eur. Phys. J. C{\bf 28}, 305
(2003); Y. Li, C. D. L\"u, J. Phys. G{\bf 29}, 2115 (2003); High
Energy Phys. Nucl. Phys. {\bf 27}, 1062(2003).


\bibitem{decayconstant}T.V. Brito, F.S. Navarra, M. Nielsen, and
M.E. Fracco, Phys. Lett. B{\bf 608}, 69(2005); K. Maltman, Phys.
Lett. B{\bf 462}, 14(1999); S. Narison, Nucl. Phys. Proc. Suppl.
{\bf 86}, 242(2000); C.M. Shakin and H.S. Wang, Phys. Rev. D{\bf
63}, 074017(2001); D.S. Du, J.W. Li and M.Z. Yang, Phys. Lett.B
{\bf 619}, 105(2005).

\bibitem{botts}J. Botts and G. Sterman, Nucl. Phys. B{\bf 325}, 62 (1989).

\bibitem{review}H.N. Li, Prog. Part. Nucl. Phys. {\bf 51},
85(2003).

\bibitem{buras}For a review, see G. Buchalla, A.J. Buras,
M.E. Lautenbacher, Rev. Mod. Phys. {\bf 68}, 1125(1996).


\bibitem{lipc}C.H. Chen, Y.Y. Keum and H.-n. Li, Phys. Rev. D{\bf 64},
112002(2003).

\bibitem{f0K}W. Wang, Y.L. Shen, Y. Li and C.D. L\"u, arXiv: hep-ph/0609082.



\bibitem{akl}A. Ali, G. Kramer, C.D. L\"u, Phys. Rev. D{\bf 58}, 094009 (1998)

\bibitem{QCDSR}V. M. Braun, I. E. Fliyakov, Z. Phys. C{\bf
48}, 239(1990); P. Ball, JHEP, {\bf 9901}, 010(1999); V.M. Braun
and A. Lenz, Phys. Rev. D{\bf 70}, 074020 (2004); P. Ball and R.
Zwicky, Phys. Lett B{ \bf 633}, 289(2006); P. Ball, V.M. Braun and
A. Lenz, JHEP {\bf 0605}, 004(2006).

\bibitem{LWZ}
C.D. L\"u, Y.M. Wang and H. Zou, hep-ph/0612210.


\bibitem{kurimoto}T. Kurimoto, Phys. Rev. D{\bf74}, 014027(2006).

\bibitem{subleadingPQCD}H.N. Li, S. Mishima and A.I. Sanda, Phys. Rev. D{\bf 72},
114005(2005); H.N. Li and S. Mishima, {\it ibid}: {\bf
73},114014(2006); arXiv: hep-ph/0608277.


\bibitem{pdg06}Particle data group, Journal of Physics G {\bf33}, 1
(2006).



\bibitem{FSI}H.Y. Cheng, C.K. Chua and A. Soni, Phys. Rev. D{\bf
71}, 014030(2005); C.D. L\"u, Y.L. Shen and W. Wang, Phys. Rev.
D{\bf73}, 034005(2006).


\end{thebibliography}
\end{document}